\documentclass[12pt]{article}
\usepackage{a4,epsfig}
\begin{document}




\begin{titlepage}

\pagenumbering{arabic}
\vspace*{-1.5cm}
\begin{tabular*}{15.cm}{lc@{\extracolsep{\fill}}r}
&
\\
& &
30 October  2017
\\
&&\\ \hline
\end{tabular*}
\vspace*{2.cm}
\begin{center}
\Large 
{\bf \boldmath
A note on a new paradox \\
in superluminal signalling     
} \\
\vspace*{2.cm}
\normalsize { 
   
   {\bf V.F. Perepelitsa}\\
   {\footnotesize ITEP, Moscow            }\\ 
}
\end{center}
\vspace{\fill}
\begin{abstract}
\noindent
The Tolman paradox is well known as a base for demonstrating the causality
violation by faster-than-light signals within special relativity. It is
constructed using a two-way exchange of faster-than-light signals between
two inertial observers who are in a relative motion receding one from another.
Recently a one-way superluminal signalling arrangement was suggested as a 
possible construction of a causal paradox. In this note we show that this 
suggestion is not correct, and no causality principle violation can occur in 
any one-way signalling by the use of faster-than light particles and signals.  

\end{abstract}
\vspace{\fill}
                    
\vspace{\fill}
\end{titlepage}




\setcounter{page}{1}    


\section{Introduction}
In 1959 E.C.G. Sudarshan has developed a hypothesis of faster-than-light
particles which was published in \cite{bds}. This publication was followed 
by a paper by G. Feinberg who coined a word $tachyon$ to name these particles
\cite{fein}.

Not surprisingly, the hypothesis of faster-than-light particles has encountered
strong objections related to the principle of causality. It has been shown 
in several papers \cite{newton,roln,benford,parment}, in agreement with 
an earlier remark by Einstein \cite{ein} (see also \cite{tolman,moller,bohm}), 
that by using tachyons as information carriers one can build a causal loop, 
making possible the information transfer to the past of an observer, thus 
creating a causal paradox. The base for the construction of casual loops is the 
so called Tolman paradox \cite{tolman,moller} in which two inertial observers, 
receding one from another, communicate via faster-than-light signals. 
This paradox, known for a while, prohibits faster-than-light particles and 
signals within Einstein's special relativity (SR) \footnote{The solution of 
this paradox in the framework of a model of a spontaneous breaking of SR is 
given in \cite{ttheor,lagran}. While conserving causality, this model 
confines the effects of the violation of the Lorentz symmetry by tachyons, 
interacting with ordinary particles, within the free tachyon sector 
(i.e. within the sector of asymptotic tachyon states), 
thus avoiding strong experimental restrictions on such violations, 
compiled e.g. in \cite{mattingly,kostel2}.}.      

Recently, in a paper by M. Fayngold, a simpler construction of a causal
paradox was suggested based on a one-way superluminal signalling \cite{oneway}.
The new paradox is based on an assumption about a non-invariance of the
information flow direction when using tachyons as information carriers.
Unfortunately, this assumption is wrong from both common sense and the 
information theory applied to superluminal communications. A correct approach 
to the problem is considered in this note. 


\section{One-way superluminal signalling}
\setcounter{equation}{0}
\renewcommand{\theequation}{2.\arabic{equation}}
Let us consider the exchange of tachyon signals between two inertial observers
(counterparts) receding one from another with the constant velocity $u$. 
In principle, under the term ``tachyon signal" one can assume a modulated 
tachyon beam carrying the arbitrarily rich information, but without loss of 
generality we adopt for the moment an approach that the signal can be 
transferred by a single tachyon.

Let us assume that the both observers are equipped with tachyon emitters, 
capable to emit fast tachyons having velocities $v > c^2/u$ to his counterpart
\footnote{At such velocities, under an appropriate Lorentz transformation, 
tachyons become antitachyons going backward in time.}, and tachyon detectors
(say, time-of-flight systems allowing to detect and identify tachyons, at least
those which go along the line connecting the observers A and B).
But first we consider a one way superluminal signalling, namely, a launching  
at $t^A_0$ of a fast tachyon $\alpha$ from the observer A to the observer B 
(Fig.~1). At the moment $t^A_1$ the tachyon $\alpha$ reaches the observer B 
and is detected by his apparatus. The sequence of these events is trivial 
from the point of view of the observer A.
 
However, it is not so trivial from the point of view of the observer B.
For him, due to Lorentz transformation, the moment $t^B_1$ precedes the 
momentum $t^B_0$, and therefore he interprets the sequence of the events as
a spontaneous emission by his detector of the tachyon signal (antitachyon 
$\overline{\alpha}$), which propagates to the observer A and is absorbed 
by the apparatus of the latter at $t^B_0$.

For some reason, this effect is considered as a violation of causality 
in many books on the SR. The reason is the orthodox formulation of 
the principle of causality: {\bf cause always precedes the effect}. 
\vspace{2cm}

\includegraphics[height=7.5cm,width=20.6cm]{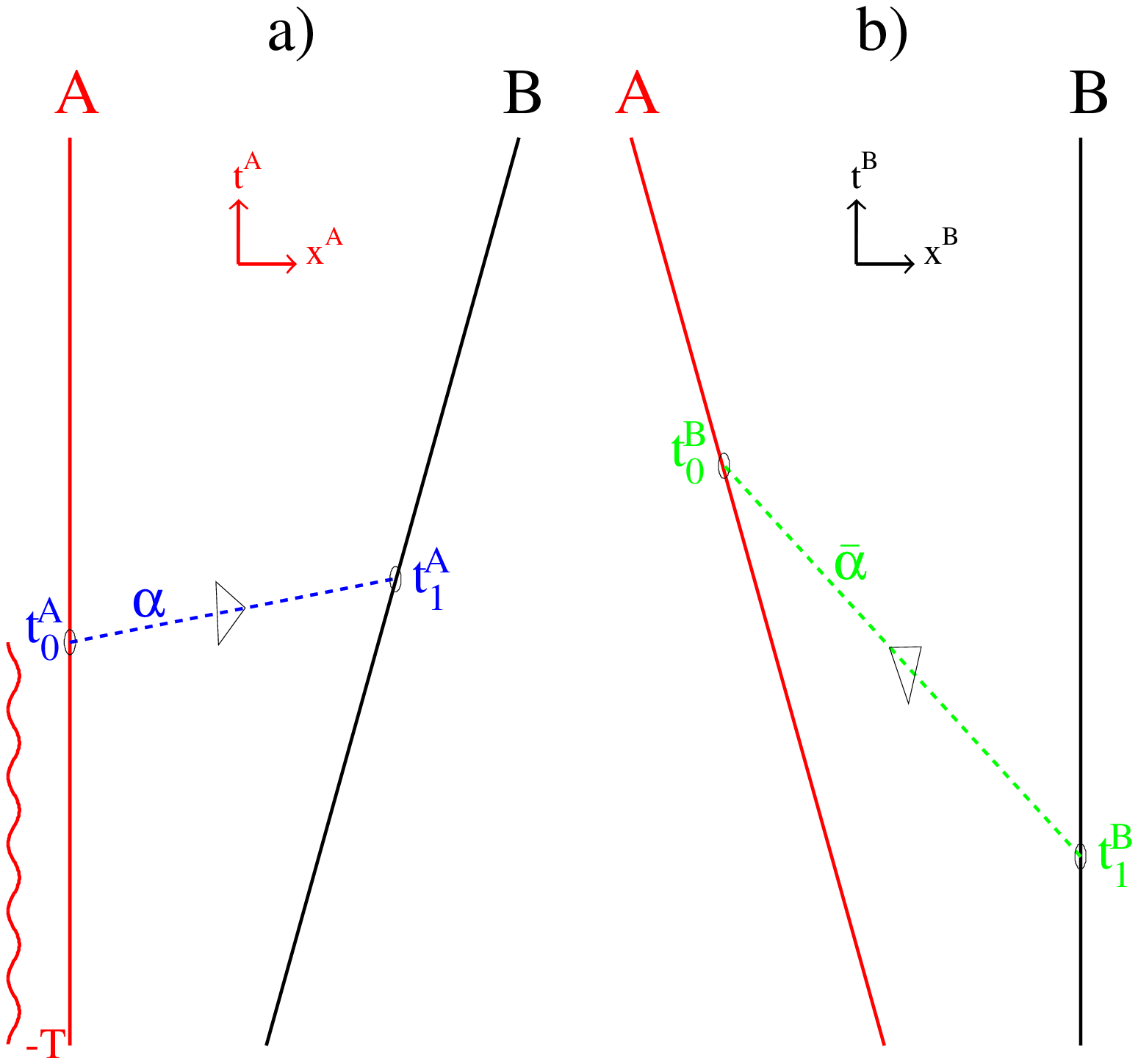}
Fig.~1. Exchange of a tachyon $\alpha$ as seen $a)$ by the observer $A$ to be
an emission of this tachyon launched by him to reach the observer $B$,
and $b)$ as seen by the observer $B$ to be a spontaneous emission of an
antitachyon $\overline{\alpha}$ moving to the observer $A$. The observer's
world lines are shown by solid lines, those for tachyons are shown by dashed
ones, the directions of tachyon motion (in space and time) being indicated 
by arrows. The ``dead time interval" preceding the emission of the tachyon
$\alpha$ at $t_0^A = 0$, symbolized by a wave line, will be used later
in the construction of the Tolman paradox.
\vspace{4mm}

Indeed, such an ordering of cause and effect allows avoiding problems with 
causality in the world of ordinary particles.
But as we shall see later, this formulation should be changed when considering 
faster-than-light particles and signals.  
\section{Construction of a one-way causal paradox and its failure}
\setcounter{equation}{0}
\renewcommand{\theequation}{3.\arabic{equation}}
\subsection{The construction}
The main idea of a construction of a one-way causal paradox suggested in
\cite{oneway} is the introduction of a third observer, C, positioned between
the observers A and B, who can control the passage of the superluminal signal 
through its position (in both directions), for example, by intercepting the 
signal by a plug which does not transmit tachyons.
Then the sequence of events is considered in both observer
frames, A and B, and an appearance of a logical inconsistency in this 
sequence is declared in the cases when the plug blocks the passage of the
tachyon signal at the position of C.

Indeed, if one follows strictly the orthodox principle of causality, as
the author of \cite{oneway} does, the introduction of the plug, viewed from the
frame A, removes the part of the tachyon world line from C to B. On the
other hand, this action, viewed from the frame of the observer B (who, 
according to \cite{oneway} with the reference to the orthodox principle of 
causality, hosts the cause in this case) should remove the part from C to A. 
Such an ambiguity is considered in \cite{oneway} as a logical paradox 
which can be used to ban faster-than-light particles and signals.

%

\subsection{The failure}
What is wrong in this construction? The answer is: it violates 
the principle of the invariance of the information flow direction.

It is easy to prove that this principle 
holds in any process of the information transfer, whatever could be 
the time order of the sending and receiving of the information (as we have 
seen, this order can be reversed in the case of the tachyon exchange). 
For the proof we may turn our consideration from a single-tachyon signal 
to the signal containing much information (e.g. transmitted by a modulated
tachyon beam). Then two straightforward arguments can be used in order to 
prove the invariance of the information flow direction.

First, we note that any 
information message presents a sequence of symbols (e.g. letters and 
numbers) separated by time-like intervals. Therefore the time ordering of the 
symbols in this sequence is invariant, i.e. it does not change even when 
the message is sent with the faster-than-light (spacelike) carriers, 
which can go backward in time.

Second, each information message can bear an identifier of its sender, so the 
source of the message can be determined without doubt in any reference frame, 
whatever could be the time order of the message sending and receiving. 

The application of the principle of the invariance of the information flow
direction results in doubtless identification of a cause and its effect in
any causally related sequence of events. Applying it to the consideration
of a concrete one-way signalling described above, we arrive to four different
situations, all of them being logically self-consistent:
\begin{itemize}
\item [a)] If the observer A issues a superluminal signal and it is blocked
by the observer C, i.e. at the position of C, all the observers record 
the pair of the events (A,~C).
\item [b)] If the observer B issues a superluminal signal and it is blocked at
the position of C, all the observers record the pair (B,~C),
\item [c)] If the observer A issues a superluminal signal and it is $not$ 
blocked by the observer C, i.e. it goes to B and is absorbed there, 
all the observers record the pair of the events (A,~B).
\item [d)] If the observer B issues a superluminal signal and it is $not$
blocked at the position of C, all the observers record the pair (B,~A),
\end{itemize} 
The time ordering inside the pairs, i.e. the time ordering of a cause, 
invariantly associated with the signal source (and presented by the first 
letter in parentheses in the items above), and its effect can be 
different for different observers, but unless a two-way signalling is
constructed (in the spirit of the Tolman paradox, described in Sect.~5)
no logical contradiction in these situations exists. 

\section{The principle of causality}
In the view of inapplicability of the orthodox principle of causality
when considering faster-than-light particles and signals, it has to be
reformulated. However, the modified principle of causality should reduce
to the orthodox one when being applied to ordinary particles.

Therefore, to get a consensus, the causality 
principle should be reformulated as follows: 
{\bf any cause has an unalterable own origin}. In other words, the causality 
principle appears to be a requirement of the impossibility of the creation of 
causal loops (i.e. causal chains containing closed world lines) which admit
the change of the conditions of their own creation 
\footnote{In particular, the modified principle of causality excludes the 
possibility of the realization of time-machine solutions of Einstein equations, 
first considered by K. G\"odel in \cite{godel}. On the other hand, it agrees
with the {\em Novikov self-consistency principle} \cite{novik}.}.
Implemented in the world of ordinary particles the modified causality principle
reproduces the orthodox one,
while in the world with faster-than-light particles it allows a consideration 
of a superluminal signalling by the exchange of tachyons, 
conserving at the same time the principle 
of the invariance of the information flow direction.

The modified principle of causality automatically excludes one-way causal
paradoxes. Referring to \cite{ttheor,lagran}, we note that accepting the 
modified causality principle one can rehabilitate the tachyon hypothesis, 
removing from it the causality problem and the problem of the tachyon vacuum 
instability, though at the price of abandoning the SR postulate about the  
equivalence of inertial frames when dealing with the faster-than-light 
particles and signals (by introducing a tachyon preferred reference frame
\footnote{An example of an originally parallel tachyon beam diverging due to
mutual interactions and/or due to interactions with other particles 
inside an isolated system (presenting a particular thermodynamic
process involving tachyons), considered first in the tachyon 
preferred reference frame and then in a moving frame in which the time sequence
of individual tachyon interactions can be reversed, shows that the concept
of entropy, associated intrinsically with the processes of the information
transfer, should share the principle of the invariance of the information
flow direction, presented in this note. The author is grateful to 
Prof.~O.~V.~Kancheli for raising the question about the increase/decrease 
of entropy of an isolated system in processes involving 
faster-than-light particles.}), 
which has to be replaced by the requirement of
a covariant formulation of a tachyon theory (from the point of view of the
present author, the price is not tremendously high). 

However, within the SR the tachyon hypothesis remains incompatible even
with the modified principle of causality, as can be demonstrated 
via the Tolman paradox construction.

\section{The Tolman paradox}
In order to make the demonstration obvious, let us introduce a mandatory
condition for the emitting of the tachyon signal $\alpha$ by the observer A:
the signal should be sent at $t^A_0$ {\em if and only if} there were no 
other tachyon signal coming from the observer B and registered by the 
apparatus of the observer A during a certain preceding period $T$. 
The duration of $T$, equal to the distance between the observers  A an B at 
the moment $t^A_0$ divided by $c$, would be sufficient for the argumentation.

As we have described above, the observer B receives this signal as coming
from his future, though from the space-like separated region, as shown in 
Fig.~1b. However, if the observer B could not produce any influence to the
signal sending (namely, to the emission of tachyon $\alpha$), i.e. the observer
$A$ would be inaccessible to the observer $B$, after detecting by the latter
the tachyonic signal $\alpha$, during a whole interval T preceding the $t_0^A$,
as it occurs in the model suggested in \cite{ttheor,lagran} in which all the
acausal tachyon states are confined within the tachyon vacuum,
then no problem with causality would appear.

Unfortunately, such a confinement is not possible within the SR. 
The observer $B$ in our example has an access to the observer $A$
during the above interval since he can possess a tachyon emitter equivalent, 
according to Lorentz symmetry, with that of the observer $A$. 
So, at the time $t_2^B$ (see Fig.~2b) he sends a faster-than-light signal 
(tachyon $\beta$) towards the observer $A$. 
If the velocity of the tachyon $\beta$ in the $B$ frame is higher
than that of the antitachyon $\overline \alpha$ the trajectory of the tachyon
$\beta$ intersects the trajectory of the antitachyon $\overline \alpha$
somewhere in the space between the $B$ and $A$, and then tachyon $\beta$ will
reach the observer $A$ and will be detected by him at the time $t_3$ which,
in the both frames, precedes the time  $t_0$. One can see that the possibility
of a causal loop is realized.
 
Let us consider this loop in the frame of the observer $A$ (Fig.~2a). He will 
detect a tachyonic signal (interpreted by him as an emission of the antitachyon 
$\overline \beta$ by his detector)
at time $t_3^A$, i.e. inside the time interval $T$ preceding 
the launching time $t_0^A$. But the mandatory condition for launching
the tachyon $\alpha$ (specified at the beginning of this Section)
was the absence of any tachyonic signal from B during
that interval. We have a capital logical paradox, which was the main reason
of a rejection of the possibility of faster-than-light signals during a
century, beginning with Einstein's formulation of this rejection \cite{ein}.
We note that often this rejection produces a mental barrier, preventing 
constructive discussion of any problem related to the faster-than-light
particles and signals, though alternative approaches to the subject also
exist. 

\vspace{2cm}
\includegraphics[height=7.5cm,width=20.6cm]{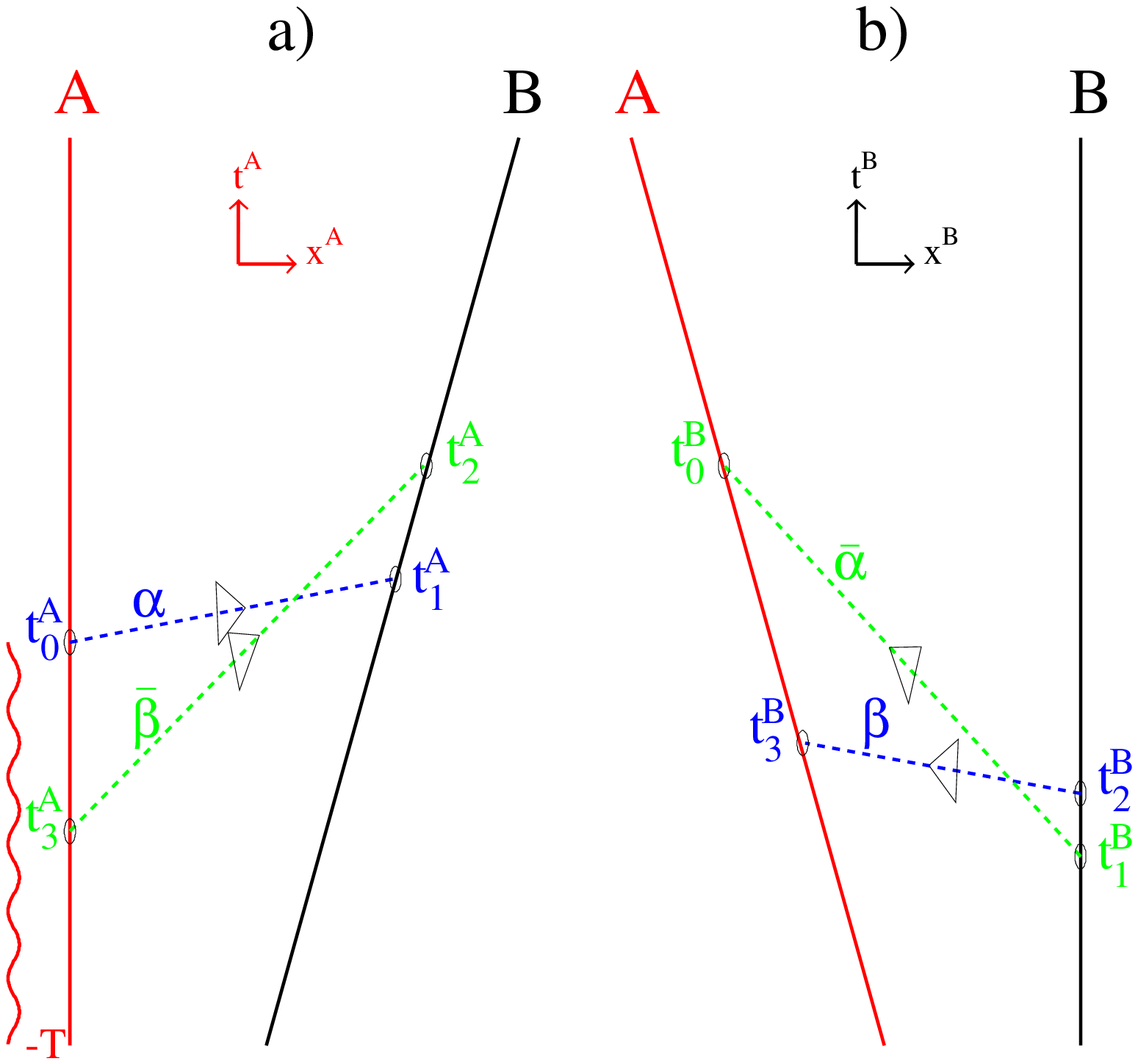}
Fig.~2. A causal loop (the Tolman paradox) as seen $a)$
in the frame of the observer $A$ and $b)$ in the frame of the observer $B$.
We emphasize once more that the arrows on the tachyon world lines indicate
the tachyon motion direction, which can be opposite (as in the cases of 
antitachyons $\overline \alpha$ and $\overline \beta$) to the information flow
direction, the latter being always directed from a cause to its effect.   

\section{Conclusion}
Turning to the concluding section of \cite{oneway} we can reformulate several
statements related to the case when the observer C blocks the tachyon signal,
not allowing a passage of tachyons through his plug,
to make these statements correct (the corrections to the
original items of \cite{oneway} are given in italic):
\begin{itemize}
\item [(a)]{\em If the observer A issues a superluminal signal} both observers 
record the pair of the events (A,~C) (albeit in the opposite ordering)
\item [(b)] {\em If the observer B issues a superluminal signal} both record 
the pair (B,~C), also in the opposite ordering
\end{itemize}  
The outcomes (c) and (d) considered in the concluding section of \cite{oneway}
should be discarded as mutually controversial.

Thus, the one-way superluminal signalling does not result in a causal paradox 
and cannot be used to ban faster-than light signals within the SR.     
Contrary, the use of a two-way construction of the Tolman paradox is crucial
in the demonstration of the incompatibility of the SR with the existence of 
tachyons due to appearance of logical inconsistencies related to the principle
of causality. This means, according to \cite{ttheor}, that this theory must not
be used when considering superluminal signalling.

\subsection*{Acknowledgements}
\vskip 3 mm
The author is grateful to Profs. K.~G.~Boreskov, F.~S.~Dzheparov, and 
O.~V.~Kancheli for useful discussions,
and to Dr.~B.~R.~French for the critical reading of the manuscript.


\end{document}